# Robust fabrication of large-area in- and out-of-plane cross-section samples of layered materials with ultramicrotomy

M. O. Cichocka[1], M. Bolhuis[1], S. E. van Heijst[1] and S. Conesa-Boj[1, *]

[1] *Kavli Institute of Nanoscience, Delft University of Technology, 2628CJ Delft, The Netherlands.*

**ABSTRACT**

Layered materials (LMs) such as graphene or $MoS_2$ have recently attracted a great deal of interest. These materials offer unique functionalities due to their structural anisotropy characterized by weak van der Waals bonds along the out-of-plane axis and covalent bonds in the in-plane direction. A central requirement to access the structural information of complex nanostructures built upon LMs is to control the relative orientation of each sample prior to their inspection *e.g.* with Transmission Electron Microscopy (TEM). However, developing sample preparation methods that result in large inspection areas and ensure full control over the sample orientation while avoiding damage during the transfer to the TEM grid is challenging. Here we demonstrate the feasibility of deploying ultramicrotomy for the preparation of LM samples in TEM analyses. We show how ultramicrotomy leads to the reproducible large-scale production of both in-plane and out-of-plane cross-sections, with bulk vertically-oriented $MoS_2$ and $WS_2$ nanosheets as proof of concept. The robustness of the prepared samples is subsequently verified by their characterization by means of both high-resolution TEM and Raman spectroscopy measurements. Our approach is fully general and should find applications for a wide range of materials as well as of techniques beyond TEM, thus paving the way to the systematic large-area mass-production of cross-sectional specimens for structural and compositional studies.

*Corresponding author: s.conesaboj@tudelft.nl

## INTRODUCTION

Recent developments in electron microscopy are one of the main drivers of the tremendous recent progress that the field of material science is experiencing. Thanks to this progress, it is now possible to provide extensive information on structural,[1] and chemical[2] properties of materials down to the single atom scale. A family of materials that has attracted a great deal of interest are layered materials[3] (LMs) such as graphene or $MoS_2$. In the context of nanostructures built upon LMs, one of the main requirements to understand their structural properties is ensuring full control over the orientation of the prepared specimens for their subsequent characterization, *e.g.* with transmission electron microscopy (TEM) measurements.

However, the preparation of samples of LMs in a manner that is suitable for their flexible inspection by means of TEM is challenging, since one needs to achieve full control over the transfer and relative orientation of these nanostructures from the substrate onto the TEM grid. For in-plane inspection, one of the most popular approaches is the direct exfoliation of LMs nanosheets from their bulk form via either mechanical or liquid methods, [4,5,6,7,8] which allows verifying their crystalline structure and chemical composition.[9,10,11] However, exfoliation limits the available inspection area since in general the totality of material cannot be transferred. In the case of out-of-plane cross-sections, these are usually prepared using focus ion beam (FIB),[12,13] which again often results in small inspections areas.

Here we adopt a popular preparation method used for biological samples,[14,15] ultramicrotomy, and demonstrate its feasibility for the preparation of LMs samples for their subsequent TEM characterization. Ultramicrotomy is being increasingly adopted as an attractive and complementary option for producing the high-quality TEM specimens,[16,17] and it can also be used to tailor the dimensions of the fabricated cross-sections.[18,19,20,21]

In this work we demonstrate the significant potential that ultramicrotomy has for the sample preparation and TEM inspection of layered materials. Our strategy achieves the reproducible large-scale production of both in-plane and out-of-plane cross-section samples of LMs, with $MoS_2$ and $WS_2$ as a proof of concept. We demonstrate how this strategy makes possible the high-resolution TEM imaging in both in the in- and out-of-plane direction. Our method is also suitable to prepare LM samples for their inspection with other techniques, as we demonstrate here by carrying out Raman spectroscopy characterization of the same cross-sections.

**RESULTS AND DISCUSSION**

Bulk vertical MoS$_2$ and WS$_2$ nanosheets grown on top of a SiO$_2$/Si substrate have been used for fabricating in-plane (with MoS$_2$) and out-of-plane (with WS$_2$) cross-sections, as indicated in **Figure 1a**. **Figures 1b** and **1c** summarize the procedures followed to create these in-plane and out-of-plane cross-sections, respectively. The two approaches differ in the orientation of the specimens within the epoxy block, which is the main factor that needs to be considered before sectioning. In the case of the in-plane cross-sections, the specimen is kept in one side of the cured epoxy block, as shown in **Figure 1b**. If our goal instead is to access the out-of-plane direction, the specimen is embedded into epoxy creating a sandwich-like assembly with the specimen in the middle of it, see **Figure 1c**. In this way, the out-of-plane direction can be aligned parallel to the knife edge used for the sectioning. In order to facilitate the access along both directions, two different embedding containers have been used, as indicated in **Figures S1** and **S2** from the supporting information (SI).

*Embedding of MoS$_2$ for in-plane cross-sections.* The embedding and curing procedures for the in-plane cross-section sample preparation is straightforward since the whole process takes place at once. The wafer was mounted on top of a cylindrical Teflon mold, see **Figure S1a**. After filling the hole with the prepared epoxy, the epoxy/wafer system was cured for 9 hours.

*Embedding of WS$_2$ for out-of-plane cross-sections.* In this particular case, a flat silicone rubber mold was used (**Figure S2a**). We began by filling half of the silicone rubber mold with the prepared epoxy. The wafer containing the specimen was placed on top of the epoxy, with the specimen side facing the epoxy, as illustrated in the left panel of **Figure 1c**. The whole epoxy/wafer system was then placed inside a furnace and cured at a temperature of 70 ºC for 4 hours. This is the minimum time required in order to make sure that the specimen can be fully detached from the silicon wafer, resulting in an epoxy/specimen system. This system was then again placed inside the silicon rubber mold, covered with epoxy, and cured for 12 hours. **Figure 1c** (right) shows a schematic representation of the final sandwich-like assembly. Notice that for these out-of-plane samples one needs to stablish an appropriate curing time for the two parts of the sandwich. This is important in order to ensure uniform thicknesses during sectioning. In addition, long curing times can efficiently prevent charging effects and contamination during the TEM inspection.

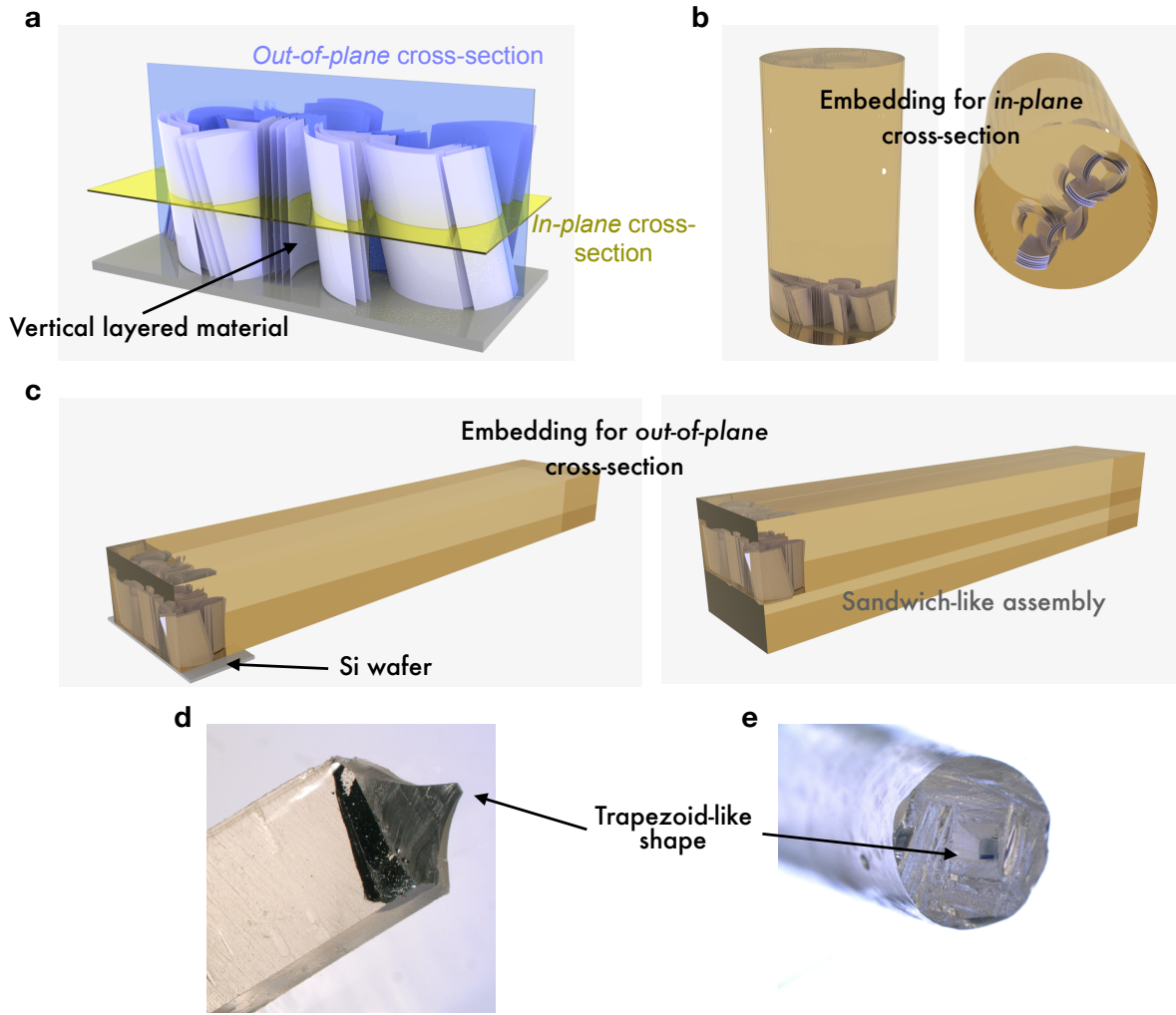

**Figure 1. (a)** Schematic representation of vertically-oriented layered materials grown on top of a substrate, where we indicate the directions corresponding to the in- and out-of-plane cross-sections. **(b)** and **(c)** illustrations representing the specimen embedded in the cured epoxy block for in- and out-of-plane configurations, respectively. **(d)** and **(e)** optical microscope images of the trimmed top of the epoxy block for out-of-plane and in-plane configurations, respectively.

*Preparing the epoxy/specimen blocks for sectioning.* Before sectioning, it is necessary to trim down the epoxy/specimen system such that its dimensions become much smaller than the length of the knife. To achieve this, we manually trimmed the side of the epoxy block containing the specimen, using both a razor blade and a scalpel. The trimmed area was reduced down to a small trapezoid-like shape with a length of around 0.5 mm, as shown in **Figures 1d** and **1e**. At the end of this procedure, the epoxy-specimen system is ready for being sectioned. For further details of the embedding, trimming, and sectioning procedures see the SI.

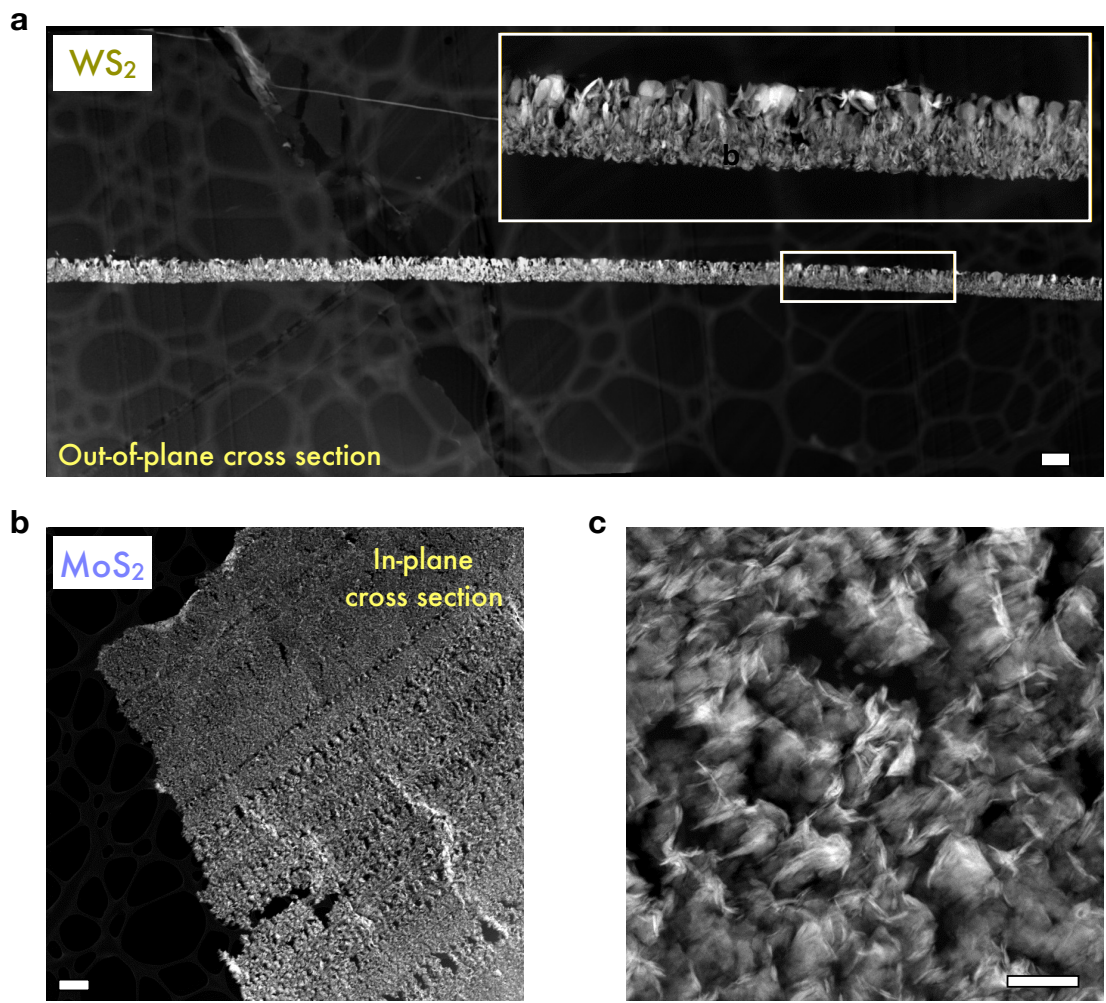

**Figure 2. (a)** and **(b)** Low-magnification HAADF-STEM images of the out–of–plane WS$_2$ and in–plane MoS$_2$ cross-sections, respectively. The inset in **(a)** displays the magnification of the rectangular area marked by a white box. **(c)** Magnified area from **(b)**. The scale bars in **(a)** and **(b)** correspond to 1 µm and in **(c)** to 200 nm.

To assess the stability of the cross-sections fabricated using the ultramicrotomy method, TEM measurements were carried out. **Figure 2** displays the low-magnification high-angle annular dark field (HAADF) scanning transmission electron microscopy (STEM) images from representative areas of the out-of-plane cross-sections of WS$_2$ (**Figure 2a**) and in-plane cross-sections of MoS$_2$ (**Figures 2b** and **2c**). From these low-magnification images, one can clearly appreciate the large cross-section areas that have been produced with our procedure for both the in-plane and the out-of-plane orientations. Remarkably, the achievable lengths of the fabricated out-of-plane cross-sections turn out to be up to several tens of microns (for a maximum length of around 35 µm). Moreover, the real advantage of the ultramicrotomy method is that these 35 µm of cross-section are contained within a single slice, and in one TEM grid it is possible to fit up to ten of these slices. This implies that the effective cross-section length available for TEM inspection is much larger, up to 350 µm, representing an improvement by more than a factor 20 as compared to the FIB results. Similar considerations apply to the in-plane $MoS_2$ cross-sections (which cannot be produced using FIB), where we achieve large effective areas suitable for TEM inspection of up to 250 µm² or even larger, depending on the density of the specimen on top of the wafer.

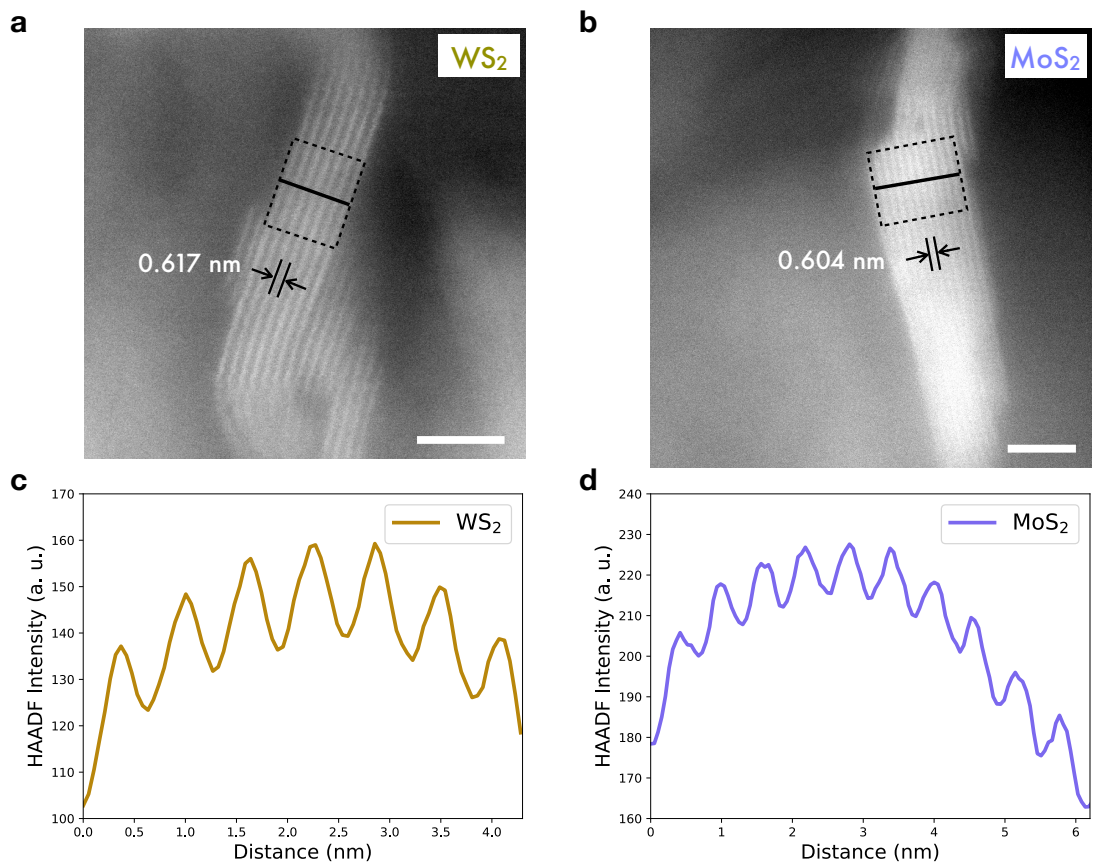

**Figure 3. (a)** and **(b)** HAADF-STEM images of vertically aligned $WS_2$ layers and $MoS_2$ layers, respectively, where the scale bars are 5 nm. In **(c)** and **(d)** The intensity profiles integrated over a rectangular area marked in **(a)** and **(b)**.

**Figures 3a** and **3b** display HAADF-STEM images of representative regions of the same out-of-plane WS$_2$ and in-plane MoS$_2$ cross-sectional specimens studied in **Figure 2**. In both cases we can resolve vertical lattice fringes. From the spacing between maxima determined from the intensity profiles integrated over a rectangular area, see **Figures 3c** and **3d**, we are able to measure the distance between these fringes. Our results indicate that the spacing between lattice fringes are 0.604±0.025 nm for MoS$_2$ and 0.617±0.025 nm for WS$_2$. These results are consistent with the (002) d-spacing within 2H-MoS$_2$ and 2H-WS$_2$ crystals.[22,23]

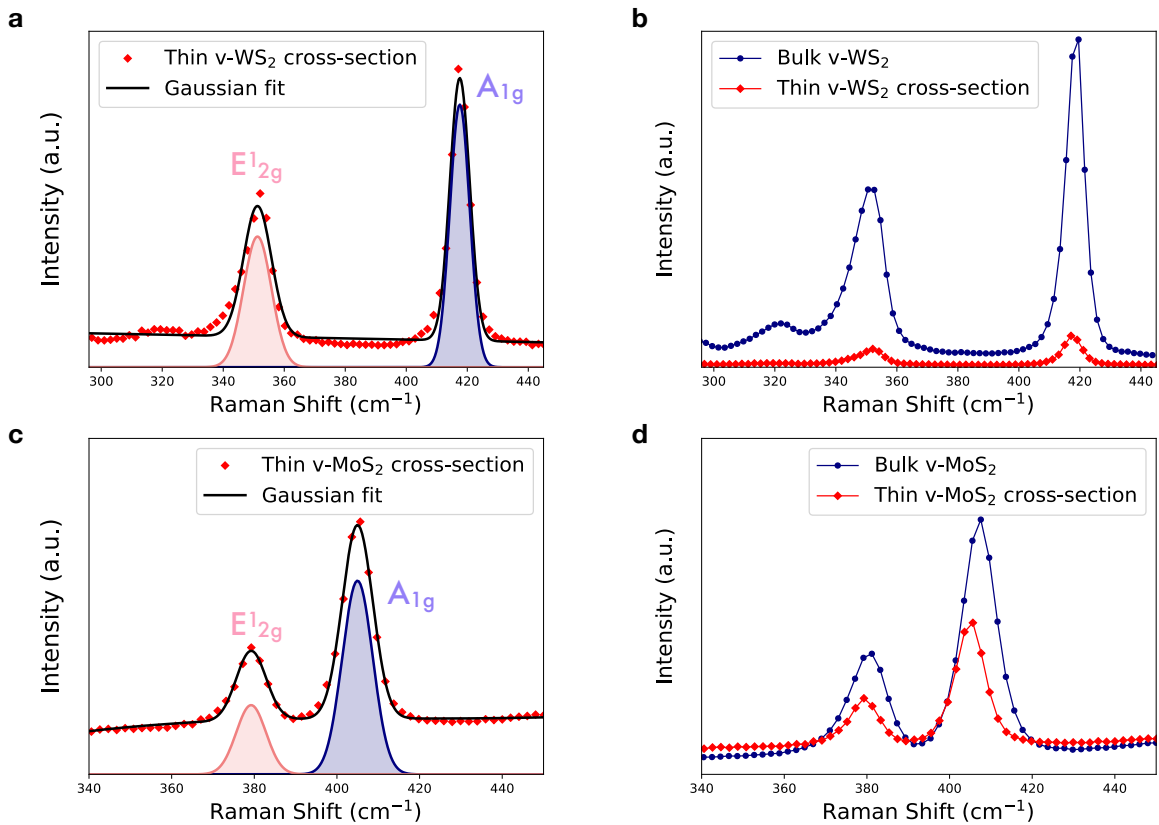

**Figure 4. (a)** and **(c)** Raman spectra obtained from the out-of-plane WS$_2$ and in the in-plane MoS$_2$ cross-sections, respectively. A Gaussian fit of the E$^1_{2g}$ and A$_{1g}$ peaks is also displayed. **(b)** and **(d)** Comparison between the Raman spectra of the out-of-plane WS$_2$ and the in-plane MoS$_2$ cross-sections with the ones obtained from the corresponding bulk vertical WS$_2$ specimen.

A crucial advantage of our approach is that it is also suitable to prepare cross-section samples of layered materials for their inspection with other techniques complementary to TEM. We have exploited this feature by carrying out room-temperature Raman spectroscopy measurements of the same WS$_2$ and MoS$_2$ cross-section samples discussed above. For these measurements we used an excitation laser with a wavelength of 514 nm, keeping the laser power under 0.5 mW at all times.

**Figure 4a** displays the Raman spectrum corresponding to the out-of-plane WS$_2$ cross-section. As one can observe, this Raman spectrum is dominated by the first-order modes, indicated by the in-plane E$^1_{2g}$ peak at 349.7 cm$^{-1}$ and the out-of-plane A$_{1g}$ peak at 417.5 cm$^{-1}$. This analysis highlights how the intensity of the out-of-plane peak (A$_{1g}$) is markedly higher than that of its in-plane counterpart (E$^1_{2g}$). Our findings are consistent with previous reports in the literature, and further confirm the vertical orientation of the WS$_2$ nanosheets. [24,25,26,27,28]

Furthermore, by comparing the Raman spectra of the out-of-plane WS$_2$ cross-section with the one obtained from the corresponding bulk specimen (**Figure 4b**) one finds that the positions of the in-plane (out-of-plane) peak shifts towards smaller (larger) frequencies by 0.7 cm$^{-1}$ (0.9 cm$^{-1}$). Is worth noting that the Raman spectrum of the bulk vertical WS$_2$ specimen was collected using the same excitation laser wavelength but with a higher laser power instead, 5 mW. For this reason, the intensity of the bulk spectrum increases significantly as compared to the cross-section case.

Moving to the Raman spectrum obtained from the in-plane MoS$_2$ cross-section sample, we find that it exhibits two main peaks located at 379.2 and 405.0 cm$^{-1}$ respectively, see **Figure 4c**. These two peaks arise from the Raman modes associated to the in-plane vibration of molybdenum and sulfur atoms (E$^1_{2g}$) and the out-of-plane vibration of sulfur atoms (A$_{1g}$). The ratio of the intensity between the two peaks, A$_{1g}$/E$^1_{2g}$, turns out to be around 3, providing further confirmation of the vertical configuration of the MoS$_2$ nanosheets.

With a similar motivation, we compare the Raman spectra corresponding to the in-plane MoS$_2$ cross-section sample with its bulk counterpart, see **Figure 4d**. From this comparison, we find that the peaks associated to the out-of-plane A$_{1g}$ and the in-plane E$^1_{2g}$ modes shift towards larger frequencies by 2.2 cm$^{-1}$ and 0.9 cm$^{-1}$, respectively.

It has been reported that the peak shifts of both the out-of-plane A$_{1g}$ and in-plane E$^1_{2g}$ modes are related to the specimen thickness variations. For instance, when comparing regular exfoliated MoS$_2$ with its bulk counterpart, one of the most relevant effects is the redshift of the A$_{1g}$ mode and blueshift of the E$^1_{2g}$ one when decreasing number of layers.[29] A similar thickness dependence is found when comparing monolayer with bulk WS$_2$ samples.[30]

In our analysis, we find that for WS$_2$ the Raman modes $A_{1g}$ and $E^1_{2g}$ shift towards larger and smaller frequencies, respectively, when increasing the specimen thickness. On the other hand, for MoS$_2$ sample both Raman peaks $A_{1g}$ and $E^1_{2g}$ undergo blue shifts upon increasing the thickness. These differences could be related to the fact that our vertical WS$_2$ and MoS$_2$ bulk specimens are scaled down in a different way for the fabrication of the out-of-plane and in-plane cross-sections.[25] In the case of the out-of-plane WS$_2$, we scale down along a single direction, similarly to what happens in the exfoliated process. While in the case of the in-plane MoS$_2$ specimen, all three spatial directions are being reduced simultaneously.

The pronounced frequency shift observed in the position of the Raman peaks, in particular for the case of the in-plane cross-section MoS$_2$, could also be an indication of structural disorder, suggesting that the local symmetry is different though the crystal structure remains the same in both cases.

These local structure variations inferred from Raman spectroscopy can be further confirmed by TEM measurements. **Figure 5** displays a reconstruction composed by low-magnification bright-field TEM images taken from the in-plane MoS$_2$ cross-section sample. We can observe that the relative orientation of the MoS$_2$ nanosheets is not homogeneous: some of them are vertically oriented as expected, but others turn out to be perpendicular to the former. This property is verified by the corresponding high-resolution TEM (HRTEM) images. As illustrated in **Figure 5b**, we can find vertically aligned nanosheets, which exhibit a hexagonal crystal system structure corresponding to the 2H-MoS$_2$ polytype. Moreover, we can also find Moiré regions arising from the superposition of different 2H-MoS$_2$ nanosheets oriented along the [001] direction and rotated with respect to each other. In **Figure 5c** we display the HRTEM image of a Moiré region in MoS$_2$, where the fast Fourier transform (FFT) in the inset illustrates the overlap of three MoS$_2$ nanosheets along the [001] direction with different relative angles. The different contributions from the three nanosheets have been indicated by different colors: red, green and blue.

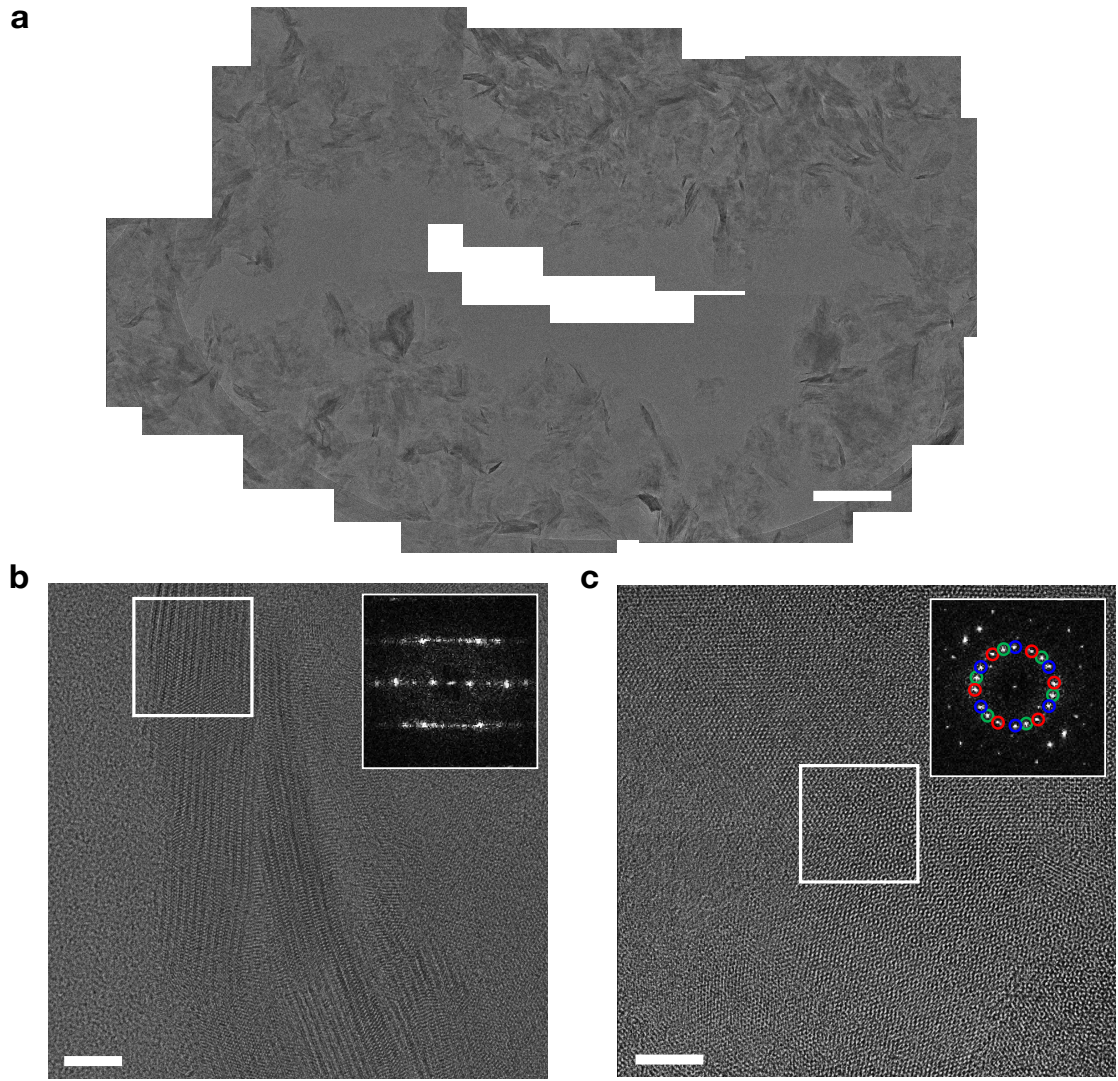

**Figure 5. (a)** Low-magnification TEM image of a $MoS_2$ patch aligned in the in-plane direction. **(b)** HRTEM image of a vertically aligned $MoS_2$ nanosheets along the [11-20] direction with its corresponding fast Fourier transform (FFT) calculated in the area marked by a white square as inset. **(c)** HRTEM image of a Moiré region in $MoS_2$, where the FFT in the inset and displays the overlap of three $MoS_2$ nanosheets along the [001] direction. The scale bar in **(a)** corresponds to 200 nm, those in **(b)** and **(d)** to 5 nm.

**CONCLUSIONS**

In this work we have presented and validated a strategy based on ultramicrotomy for the efficient and reproducible fabrication of large-area in-plane and out-of-plane cross-sections of layered materials, with $MoS_2$ and $WS_2$ as proof of concept. This approach, that bypasses some of the limitations that affect other cross-section sample preparation techniques, allows one to reliably manipulate and control the relative orientation of the prepared samples for their characterization without affecting or damaging the specimen in any way.

We have verified the structural robustness of the resulting cross-section specimens by means of detailed characterization studies using both Transmission Electron Microscopy and Raman spectroscopy measurements. Remarkably, the structural insights provided from the TEM and the Raman data confirm each other findings and demonstrate that our ultramicrotomy-based approach can be reliably adopted for the systematic structural characterization of in-plane and out-of-plane LM specimens.

While in this work we have demonstrated the viability of the ultramicrotomy-based cross-section sample preparation approach for the case of $MoS_2$ and $WS_2$, the method is fully general and should find applications enabling novel structural studies for several other layered materials, including for example topological insulator materials. As such, we believe that the method proposed here should represent an important contribution to the ongoing efforts of the nanotechnology community, paving the way to the systematic large-area mass-fabrication of cross-sectional specimens for advancing the structural and compositional inspection in a wide range of materials.

**EXPERIMENTAL SECTION**

**Layered material synthesis.** As a proof of concept of our proposed microtomy-based strategy for sample preparation, we have carried out the sectioning of two different LMs, namely bulk $MoS_2$ and $WS_2$ vertical nanosheets. These were synthesized using a two-step process, where a 700 nm-thick Mo (186 nm-thick W) layer was first sputtered onto $SiO_2$/Si substrate, and subsequently sulphurised at 700 ºC (750 ºC) into $MoS_2$ ($WS_2$). The sulfurization process was carried out in a gradient tube furnace from Carbolite Gero. The sample was placed in the middle zone and gradually heated to 700 ºC (750 ºC) at a rate of 10 ºC/min. Once the sample reached the desired temperature, 400 mg of sulfur was heated to 220 ºC. The sulfur was placed upstream from the sample. Note that Argon gas was used as a carrier gas.

In the following, we present the details of the three main steps required for the sectioning procedure: embedding into epoxy, trimming, and finally sectioning using ultramicrotomy.

**Embedding into epoxy.** All the specimens were embedded in Agar Low Viscosity resin, from Agar Scientific. The ratio between the hardeners (VH1 and VH2) and the curing temperature was optimized to achieve the desired hardness, while the two other components of the recipe, the low viscosity resin and the accelerator, were kept fixed at values of 48g and 2.5g. For further details, see **Table S1** of the supporting information (SI).

**Ultramicrotomy.** The trimming and sectioning have been carried out using an RMC Ultramicrotome PowerTome PC from Boeckeler Instruments. For the trimming of the cured epoxy block containing the specimen, a razor blade and a sculpt were applied. For the sectioning procedure, we employed a DiATOME diamond knife with a cutting angle of 35º and a clearance angle at 6º. The cutting speed and the sectioning thickness were generally set to 1mm/s and a programmed thickness of 20 or 30 nm respectively. The ribbons of the slices were directly collected from the deionizer bath and deposited onto a 300-mesh lacey carbon film copper TEM grid. The ribbons have been dried using a filter paper when deposited onto the TEM grid.

**Characterization techniques**. Transmission Electron Microscopy (TEM) measurements were carried out in a Titan Cube microscope using an acceleration voltage of 300 kV. Its spatial resolution at Scherzer defocus conditions is 0.08 nm in the High-Resolution Transmission Electron Microscopy (HRTEM) mode, whilst the resolution is around 0.19 nm in the HAADF-STEM (High-Angle Annular Dark Field – Scanning Transmission Electron Microscopy) mode.

**Raman spectroscopy.** The Raman spectroscopy measurements were carried out using a Renishaw Invia Reflex Microscope. A 514 nm incident laser beam was used in a backscattering configuration. The signal was analyzed with a 1800l/mm grating, resulting in a spectral resolution of around 1 cm$^{-1}$.

**ACKNOWLEGMENTS**

M.O.C. acknowledges support from the Netherlands Organizational for Scientific Research (NWO) through the Nanofront program. M.B., S.E.v.H., and S.C.-B. acknowledge financial support from ERC through the Starting Grant "TESLA" grant agreement No. 805021.

**Author Contributions**

M.B and S.E.v.H. synthesized the vertical $MoS_2$ and $WS_2$ samples, respectively. M.O.C performed the Ultramicrotomy and the TEM measurements. M.O.C analyzed the TEM data. M.B and S.E.v.H. performed and analyzed the Raman measurements. M.O.C, M.B, S.E.v.H. and S.C.-B prepared the figures and the discussion of the results. S.C.-B designed and supervised the experiments. All the authors contributed to the writing of the manuscript.

**Corresponding author**

Correspondence about this work should be send to Sonia Conesa-Boj.

Supporting Information

# Robust fabrication of large-area in- and out-of-plane cross-section samples of layered materials with ultramicrotomy

*Magdalena O. Cichocka[1], Maarten Bolhuis[1], Sabrya E. van Heijst[1] and Sonia Conesa-Boj[1,\*]*

[1] Kavli Institute of Nanoscience, Delft University of Technology, 2628CJ Delft, The Netherlands

*E-mail: s.conesaboj@tudelft.nl

**Section I: Embedding, trimming and sectioning.**

***Embedding into epoxy.*** All the specimens discussed in this work were embedded in an Agar Low Viscosity resin, from Agar Scientific. The ratio between the two hardeners, VH1 and VH2, as well as the curing temperature was optimized in order to achieve the desired hardness of the epoxy blocks. The two other components of the recipe, namely the amount of low viscosity resin and of the accelerator, were kept fixed at their baseline values of 48 g and 2.5 g respectively. **Table S1** summarizes the different amounts of the components used for the premixed embedding epoxy mixture, from which their relative proportion can be determined.

The resulting epoxy mixture was further optimised by means of adjusting the curing temperature and the curing times. We found that a curing temperature between 60 ºC and 70 ºC together with around 16 and 9 hours of curing time represent reasonable parameters to ensure a suitable polymerization of the resin, since these choices result in epoxy blocks with consistent and reproducible properties.

| Component | Structure | Weight (g) |
|---|---|---|
| Low Viscosity Resin | - | 48 |
| Low Viscosity Hardener (VH1) | (2-Nonen-1-yl) Succininc anyhrdide | 12 |
| Low Viscosity Hardener (VH2) | 1,2,3,6-tetrahydromethyl-3,6-methanophthalicanhydride, MNA | 40 |
| Accelerator | Benzyldimethylamine, BDMA | 2.5 |

**Table S1.** Components of the premixed embedding epoxy mixture. In each case we indicate its structure and the amount used for the optimal recipe used here.

The embedding procedure used to create the in-plane (for $MoS_2$) and out-of-plane (for $WS_2$) cross-sections differ in the relative orientation of the specimens within the epoxy block. As discussed in the main text (see **Figure 1**), this orientation is the main factor that needs to be considered before sectioning. In the case of in-plane cross-sections the embedding procedure is straightforward given that the whole process takes place at once. We used a cylindrical Teflon mold (**Figure S1a**) as a container for preparing the epoxy block. First, the wafer was placed below the cylindrical container, aligned with the hole. Afterwards, the hole was filled with the premixed epoxy. Then the wafer/epoxy/container system was placed inside a furnace where the epoxy was cured for 9 hours. **Figure S1b** shows the cured epoxy block containing the wafer once it has been removed from the cylindrical container. Such a curing time allowed detaching the wafer while maintaining the specimen inside the epoxy block (**Figure S1c**).

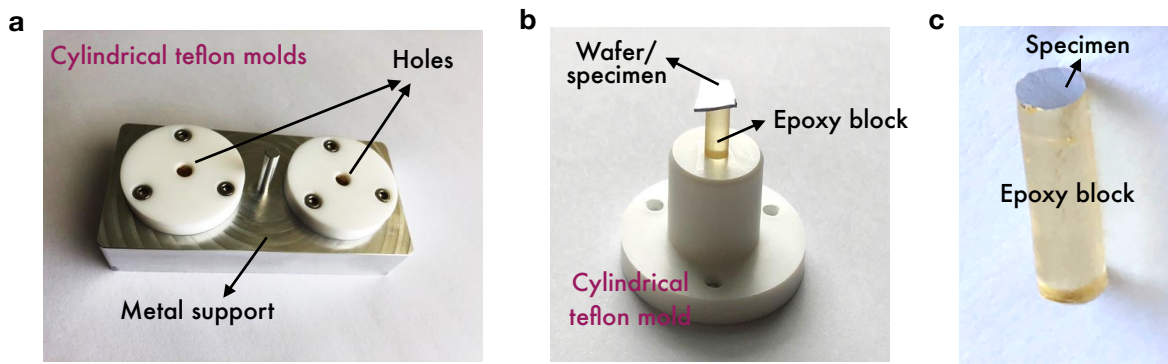

**Figure S1. (a)** The cylindrical Teflon mould used as container for the preparation of the epoxy blocks. **(b)** The cured epoxy block containing the wafer with the specimen and **(c)** the resulting epoxy block once we have detached the wafer containing the specimen.

Concerning the preparation of the epoxy blocks for the out-of-plane ($WS_2$) cross-sections, a two-step process was followed. Here a flat silicone rubber mold was used (**Figure S2a**). To begin with, half of the silicone mold was half-filled with the premixed epoxy. The wafer/specimen was then placed on top the epoxy, with the specimen side facing the epoxy (**Figure S2b**). The whole epoxy/wafer system was then placed inside a furnace and cured at a temperature of 70 °C during 4 hours. We determined that this was the minimum amount of time required to ensure that the wafer/specimen could be detached from the half-epoxy block. This epoxy/specimen was again placed inside the silicone rubber mold and covered with epoxy and cured for a further 12 hours. **Figure S2c** displays the final sandwich-like assembly which we then use for the fabrication of out-of-plane cross-sections.

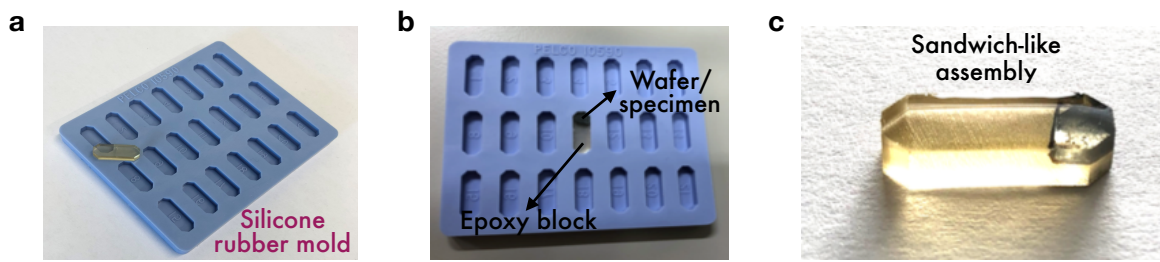

**Figure S2. (a)** The flat silicone rubber mould used to fabricate the out-of-plane WS$_2$ cross-sections. **(b)** The wafer/specimen on top of the half-filled epoxy in the silicone mold. **(c)** The final resulting sandwich-like assembly.

It is worth emphasizing that during the filling process it is crucial to reduce or ideally to avoid altogether the possible incorporation of air bubbles in the mixture. We found that by keeping the prepared epoxy mixture at a room temperature for between 10 and 15 minutes it was possible to reduce significantly the number of air bubbles formed during the preparation of the premixed epoxy.

*Trimming and sectioning.* The trimming and sectioning have been carried out using an RMC Ultramicrotome PowerTome PC from Boeckeler Instruments (**Figure S3a**). For the trimming of the cured epoxy block containing the specimen, a razor blade and a sculpt were used. Here it is very important to select the trimmed area accordingly to the orientation of the embedded specimen in the epoxy block. The obtained trapezoid-like shape face has a size of around 0.5 mm, with its area being 0.290 mm × 0.228 mm and 0.488 mm × 0.563 mm for the resulting out-of-plane and in-plane orientations respectively, see **Figures S3b** and **S3d**.

For the sectioning procedure, we employed a DiATOME diamond knife with a cutting angle of 35º and a clearance angle at 6º. The cutting speed and the sectioning thickness were generally set to 1 mm/s and a programmed thickness of 20 or 30 nm was adopted respectively. These ribbons of the slices were collected from the deionizer bath with a perfect loop and subsequently deposited onto a 300-mesh lacey carbon film copper TEM grid. These ribbons were then dried using a filter paper deposited onto the TEM grid (**Figures S3c** and **S3e**).

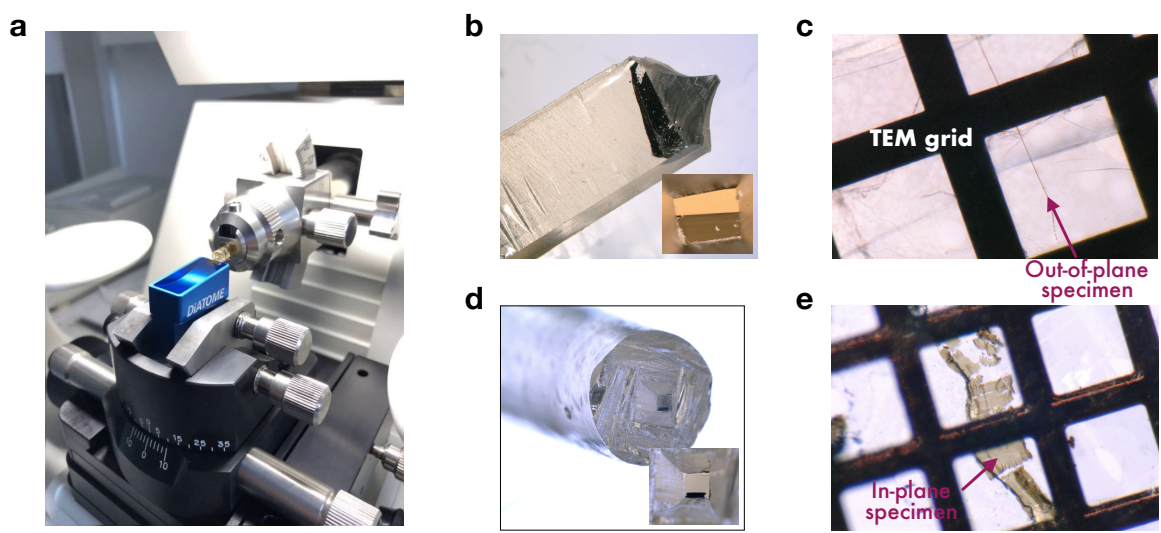

**Figure S3. (a)** The RMC Ultramicrotome used for the trimming and sectioning of our samples. **(b)** and **(d)** Resulting trapezoid-like shapes after the trimming of the cured epoxy block for the out-of-plane and in-plane specimens, respectively. **(c)** and **(e)** Optical microscope images of the TEM grids with the ribbons containing the specimens on top.